\documentclass[aps,prb,superscriptaddress,twocolumn]{revtex4-1}
\usepackage{amsmath}
\usepackage{color}
\usepackage{floatrow}
\usepackage{amssymb}
\usepackage{graphicx}
\usepackage{xcolor}
\usepackage{blindtext}
\usepackage{multirow}
\usepackage{booktabs}
\usepackage{chngcntr}
\usepackage{subfigure}

	\def\etal{{\it et al. }}

\begin{document}

\title  {Electronic and optical properties of Germagraphene, a direct band-gap semiconductor.}
\author{Sujoy Datta}
\email{sujoydatta13@gmail.com}
\affiliation{Department of Physics, University of Calcutta, Kolkata 700009, India}
\affiliation{Department of Physics, Lady Brabourne College, Kolkata 700017, India}
\author{Debnarayan Jana}
\affiliation{Department of Physics, University of Calcutta, Kolkata 700009, India}
\author{Chhanda B. Chaudhuri}
\affiliation{Department of Physics, Lady Brabourne College, Kolkata 700017, India}
\author{Abhijit Mookerjee}
\email{abhijit.mookerjee61@gmail.com}
\affiliation{(Retired Professor Emeritus) S. N. Bose National Centre for Basic Sciences,\\ Salt Lake City, Kolkata 700098, India}

\begin{abstract}
In this communication, we report a theoretical attempt to understand the electronic and optical properties of germagraphene, a two-dimensional graphene analogue. We study two different structures, C$_{17}$Ge and C$_{16}$Ge. In the C$_{17}$Ge structure, a germanium atom replaces a carbon atom while in C$_{16}$Ge structure, a carbon-carbon bond is replaced by a single germanium atom. These two types of doping have been experimentally made possible by Tripathi \etal [{\it{ACS Nano (2018) 1254641-4647}}]. We find that C$_{16}$Ge has a planar structure, whereas, the Ge atom in C$_{17}$Ge settles in an out-of-the plane position, resulting in a buckled structure. Due to Ge doping, the band-gaps open up in both. The 1.227 eV direct gap of C$_{17}$Ge is ideal for effective light absorbance and optoelectronic devices. Further study of optical properties supports this claim as well. 
\end{abstract}

\date{\today}
\maketitle

\section{Introduction}
   The discovery of graphene \cite{graphene} and related compounds revolutionized modern semiconductor industry. Subsequently, other 2D materials like silicine \cite{silicine}, black-phosphorene \cite{phosphorene}, and borophene \cite{borophene} were synthesized experimentally. 
 
 Though graphene exhibits extraordinary thermal, mechanical and electrical properties, its zero band-gap poses a severe constraint on its practical applicability. As a result, opening up  the band-gap of graphene and other graphenic zero-gap materials has been considered to be the  top-most priority in semiconductor engineering. It has been seen earlier that introducing defects in graphene or graphene-nanotubes has a large influence on their electronic properties \cite{dope1, dope2, dope3, das2016, nanotube_dope, nanotube_dope2, nanotube_sujoy}. 

Experimental work on such materials have been prolific. The aim of this communication is to use the latest theoretical approaches, together with what we shall argue to be  better modifications, and explain these experimental results. Dependable theoretical predictions will then provide the experimentalist a much better handle in choosing their desired materials out of a plethora of possibilities.

  Silicon and Germanium being the same group IV material as Carbon, are always the first choice for doping graphene. Several theoretical studies on silicine \cite{chowdhury2016} and silicon doped graphene, i.e., siligraphene \cite{siligraphene1, siligraphene2, siligraphene3} have been carried out by varying the relative concentrations of Silicon and Carbon. Siligraphene with $Si:C=1:7$ showed superior sunlight absorbance. \cite{siligraphene1} Wang \etal found theoretically that siligraphene had wonderful Li-ion storage capacity\cite{siligraphene1}. Theoretically Ge doping was also seen to tune the band-gap\cite{germagraphene-bg, germagraphene-bg}. Very recently Tripathi \etal successfully implanted Ge on graphene\cite{expt-germagraphene}. Hu \etal theoretically studied lithium ion absorption in germagraphene and found C$_{17}$Ge to be  stable in both studies. \cite{germagraphene-lithium}

  Chemical doping often results in breaking the symmetry of the system lattice. Ge is a heavier atom than C. When doped with C results in off-the-plain buckling. Such buckling has been reported in earlier theoretical investigations as well. \cite{germagraphene-lithium} However, when Ge replaces a C-C bond, it can be accommodated in the plane. We shall theoretically study both of these types of substitutional defects.

 Dong \etal have reported that the stable siligraphene structure (SiC$_7$) exhibited a direct band-gap and superior light absorbance making it a promising donor material for optical-devices\cite{siligraphene1}. So, there is a pressing need to theoretically  explore the optical properties of stable germagraphene structures.

\section{Computational Details}

The basic calculations were carried out in plane wave based techniques used in the Quantum Espresso (QE) codes\cite{QE1, QE2}. Slab geometry was simulated by introducing 12 \AA ~vacuum on  either side of the slabs. All structures were optimized first. Variable cell structural relaxations were done using the projected augmented wave method (PAW) of the Quantum Espresso (QE) using Perdew-Burke-Ernzerhof (PBE) exchange-correlation potentials\cite{PBE,PBE1}. Charge-densities and energies for each calculation were converged to 10$^{-7}$ with the maximum force of 0.001 Ry./atom.  6$\times$6$\times$6 k-point meshes were used and force convergence and pressure thresholds were set as 0.0001 Ry/au and 0.5 Kbar, respectively.

It is well-known that both the Local Density Approximation (LDA) and the PBE based Generalized Gradient Approximation (PBE-GGA) underestimate  band-gaps. \cite{pplb, singh2013} So, as a first improvement, we introduced the  Heyd$-$Scuseria$-$Ernzerhof (HSE) screened hybrid functional method\cite{HSEtheo, HSEtheo1} for electronic structure calculations. 

To get a clear idea of the origin of the gaps, we extracted the Wannier orbitals \cite{wannier} from QE. Those orbitals were maximally localized using wannier90.\cite{wannier90}

For optical property predictions, we calculated the complex dielectric tensor using HSE. Random phase approximation (RPA) was used to extract the complex dielectric tensor.
\begin{eqnarray}
\epsilon_{\alpha\beta}(\omega)&= 1+\frac{4 \pi e^2}{\Omega N_{\textbf{k}} m^2}\sum\limits_{n,n'}\sum\limits_{\textbf{k}}
\frac{|\langle u_{\textbf{k},n'}\vert\hat{\textbf{p}}_{\alpha}
\vert u_{\textbf{k},n}\rangle|^2}{(E_{\textbf{k},n'}-E_{\textbf{k},n})^2} \nonumber\\
&\left[\frac{f(E_{\textbf{k},n})}{E_{\textbf{k},n'}-E_{\textbf{k},n}+\hbar\omega+i\hbar\Gamma} +
\frac{f(E_{\textbf{k},n})}{E_{\textbf{k},n'}-E_{\textbf{k},n}-\hbar\omega-i\hbar\Gamma}\right] \nonumber\\
& \phantom{jk}
\end{eqnarray}

Here, $\Gamma$ is the inter-smearing term tending to zero. Since no excited-state can have infinite lifetime, we have introduced small positive $\Gamma$ in order to produce an intrinsic broadening to all exited states. The imaginary part of the dielectric function $\epsilon^{(i)}_{\alpha\beta}$ had been calculated first and the real part $\epsilon^{(r)}_{\alpha\beta}$ was found using the Kramers-Kronig relation. 

\[ \epsilon^{(r)}_{\alpha \beta}(\omega)=1+\frac{2}{\pi}\int_{0}^{\infty}\frac{\omega' \epsilon^{(i)}_{\alpha \beta}(\omega')}
{\omega'^{2}-\omega^{2}}d\omega'
\] 

 Optical-conductivity, refractive index and absorption-coefficients were calculated using real and imaginary parts of dielectric functions.\cite{wooten}

\begin{eqnarray}
&\text{Dielectric tensor:  } \epsilon_{\alpha\beta}=\epsilon_{\alpha\beta}^{(r)}+ i \epsilon^{(i)}_{\alpha\beta} \\ 
&\text{Optical Conductivity:  } Re [\sigma_{\alpha\beta} (\omega)]= \frac{\omega}{4\pi}\epsilon^{(i)}_{\alpha\beta}(\omega)
\end{eqnarray}

\begin{eqnarray}
&\text{Complex Refractive Index:  } \mu_{\alpha\alpha}=n_{\alpha\alpha}+i k_{\alpha\alpha}\\ \nonumber
&n_{\alpha\alpha}(\omega)= \sqrt{\frac{|\epsilon_{\alpha\alpha}(\omega)|+\epsilon^{(r)}_{\alpha\alpha}(\omega)}{2}};
k_{\alpha\alpha}(\omega)= \sqrt{\frac{|\epsilon_{\alpha\alpha}(\omega)|-\epsilon^{(r)}_{\alpha\alpha}(\omega)}{2}}
\end{eqnarray}
%\end{flushleft}
\begin{equation}
\text{Absorption Coefficient: } A_{\alpha\alpha}(\omega)=\frac{2\omega k_{\alpha\alpha}(\omega)}{c}
\end{equation}

\section{Results and discussion}

\subsection{C$_{17}$Ge}

{\bf Crystal Structure:}\hfill\\

\begin{figure}[h!]
	\centering	
	\framebox{\includegraphics[scale=0.033]{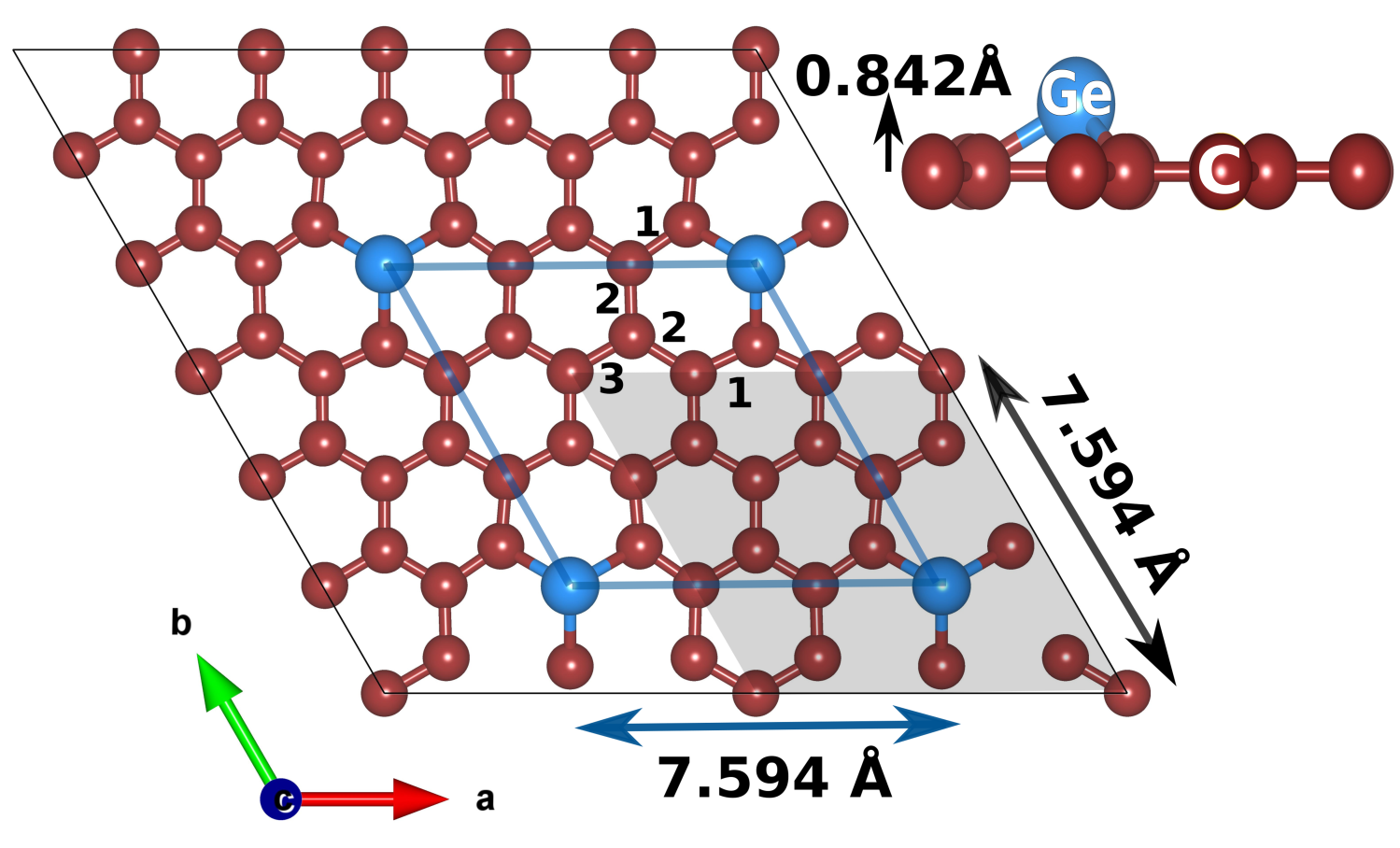}}
	\caption{Crystal structure of C$_{17}$Ge. Optimized bond lengths labelled by 1, 2, 3 are 1.4125\AA, 1.4765\AA and 1.4581 \AA, while the C-Ge bond-length is 1.8388\AA. C-Ge-C, C-C-Ge and C-C-C bond angles are 100.69$^o$, 112.314$^o$ and 123.319$^o$.}
	\label{fig1}
\end{figure}

A 3$\times$3 supercell using the unit cell of graphene (P6/mmm symmetry group) was built first (shown as shaded region in Fig. \ref{fig1}). Then a C atom was replaced by Ge and the structure was relaxed.

 The geometrically relaxed structure has the Ge atom at 0.842\AA ~off-the-plane. The in-plane lattice constant is 7.594\AA ~and the separation between layers is kept at 12\AA ~to nullify and intra-layer interaction, prerequisite for 2D calculations. We can clearly visualize a rhombic formation of Ge atoms. C-Ge bond length is 1.8388\AA, whereas, the C-C bonds denoted by 1, 2, 3 in Fig.\ref{fig1} have lengths of  1.4125, 1.4765, 1.4581 \AA ~ respectively. So, all the hexagonal rings with only carbon atoms are not regular-hexagons after Ge doping.C-Ge-C, C-C-Ge and C-C-C bond angles are 100.69$^o$, 112.314$^o$ and 123.319$^o$, respectively. Being a heavier atom, Ge tends to distort the structure more than that of siligraphene structures. \cite{siligraphene1}. The buckling structure suggests a deviation from the pure sp$^2$ hybridization which is a signature of planar structure.

\begin{figure}[b]
	\centering
		\framebox{\includegraphics[scale=0.25]{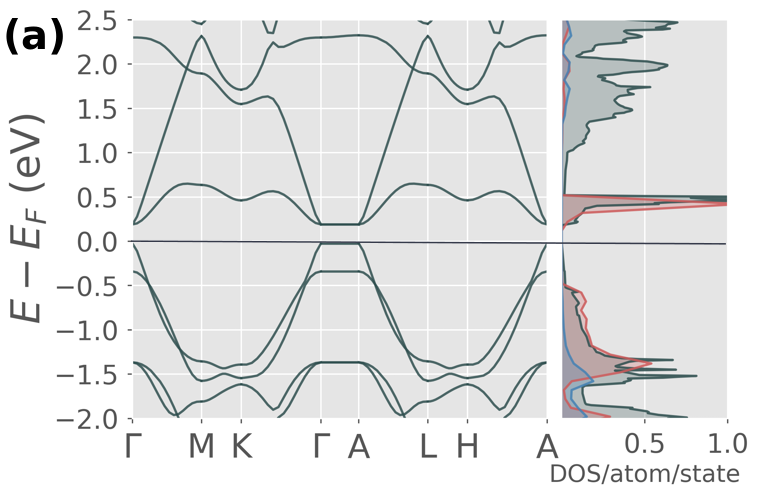}}
		\framebox{\includegraphics[scale=0.071]{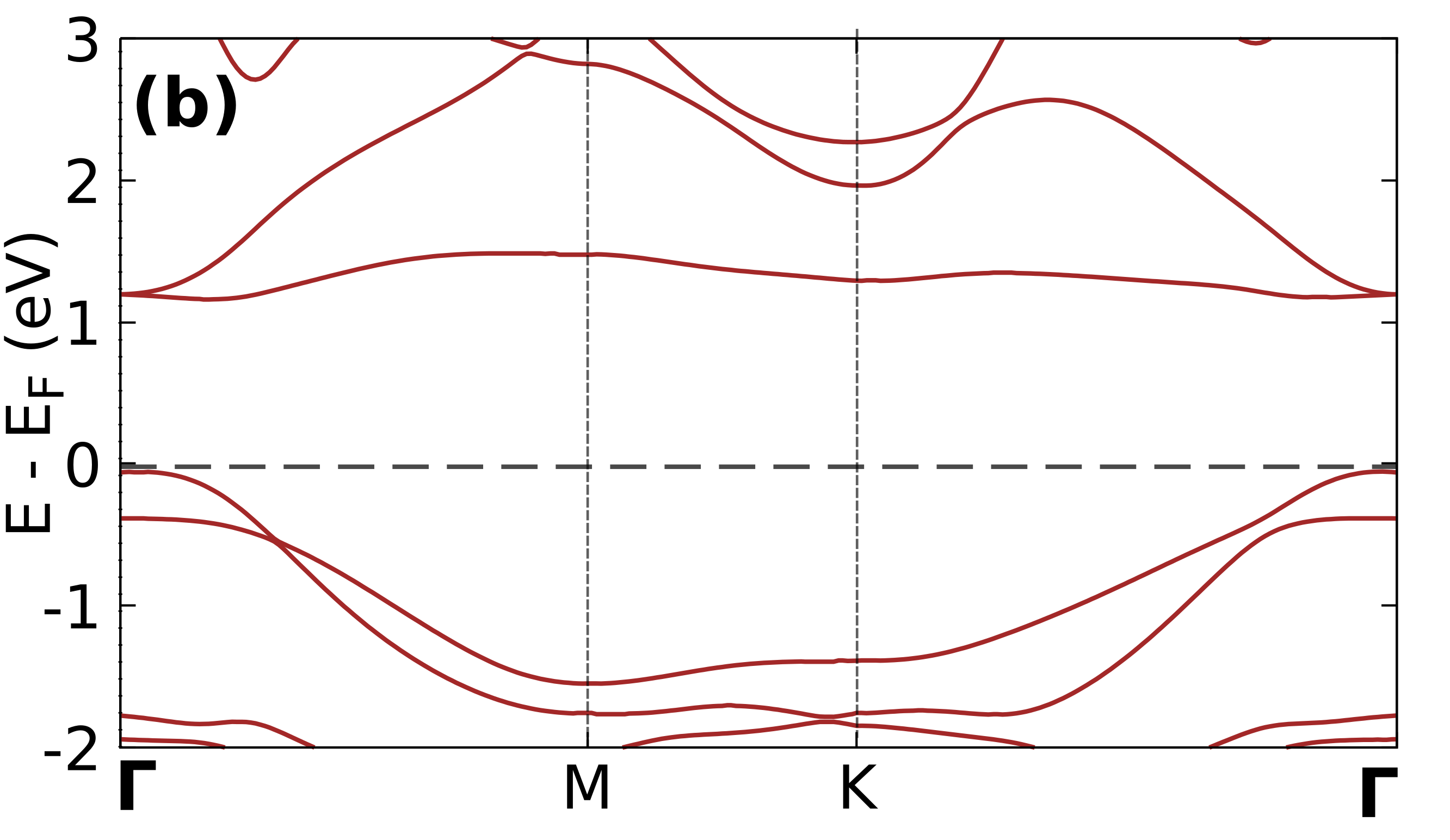}}
	\caption{(a)The band structures and density of states for C$_{17}$Ge are shown here. The results in (a) was calculated using PBE and in (b) was calculated using HSE. The p$_z$ and p$_x$ projected partial DOS of Ge are shown in red and blue, whereas, the total DOS is in grey. Note that the PBE gap is 0.2186 eV as compared with the HSE gap is 1.2271 eV.}  
	\label{fig2}
\end{figure}

{\bf Electronic Properties :}\hfill\\

The band structure and densities of states (DOS) for C$_{17}$Ge are plotted in Fig.\ref{fig2}. The reference energy is taken to be the Fermi energy. From band structure plot it is evident that C$_{17}$Ge is a direct band-gap semiconductor. The estimated band-gap using PBE is 0.2186 eV at $\Gamma$ point. The total DOS/ atom / state is plotted in grey. Approaching E$_F$ from the occupied states below, the total DOS (TDOS) shows a gradual fall, whereas, there is a sharp peak at the lowest unoccupied (LU) state. To understand the origin of these states, we plot the p$_z$ and p$_x$ projected DOS of Ge in red and blue, respectively. As there is no Ge states in between $-0.3eV$ and $0eV$, so, we can conclude that the band starting at $-0.3eV$ at $\Gamma$ is the highest occupied band originating from Ge. The HO state belongs to carbon atom. The sharp red peak at LU state indicates that the LU state belongs to Ge. So, the resulting gap is between HO of C and LU of Ge. This confirms the effect of doping in band gap opening.

Due to derivative discontinuity , LDA and PBE-GGA often underestimate the band-gap. The origin of this derivative discontinuity has been well explored in literature. \cite{pplb, harbola, singh2013, sujoy2019} We performed HSE hybrid functional calculation for band-gap prediction. The HSE calculated gap is 1.2271 eV. So, there is a derivative discontinuity factor of $\sim 1eV$ in the calculation of band-gap.
 
To plot the band-structure using HSE, we extracted the information of Wannier orbitals from QE and then localized those orbitals in wannier90 package. This helps us to understand the origin of the bands better. In Fig.\ref{fig2}(b), the plot for most localized p$_Z$ orbitals are shown. As we already predicted, the gap is between the p$_Z$ states, is verivied using this plot.
 
Previous calculation using Tran-Blaha modified Becke-Johnson exchange-correlation potential has shown 0.726 eV gap for in-plane Ge doped graphene structure at same composition\cite{germagraphene-mBJ}. Being  structures of different symmetry, we cannot really compare the results though doping percentage is the same. It should be noted that phononic stability of this C$_{17}$Ge structure has already been reported by Hu \etal.\cite{germagraphene-lithium}. Direct band-gap semiconductor is always preferable for optoelectronic and transistor devices\cite{opto}. For light absorbance, too, direct gap semiconductors performs better.

\subsection{C$_{16}$Ge}

 {\bf Crystal Structure: }\hfill\\

\begin{figure}[h]
	\centering
	\framebox{
		\includegraphics[scale=0.04]{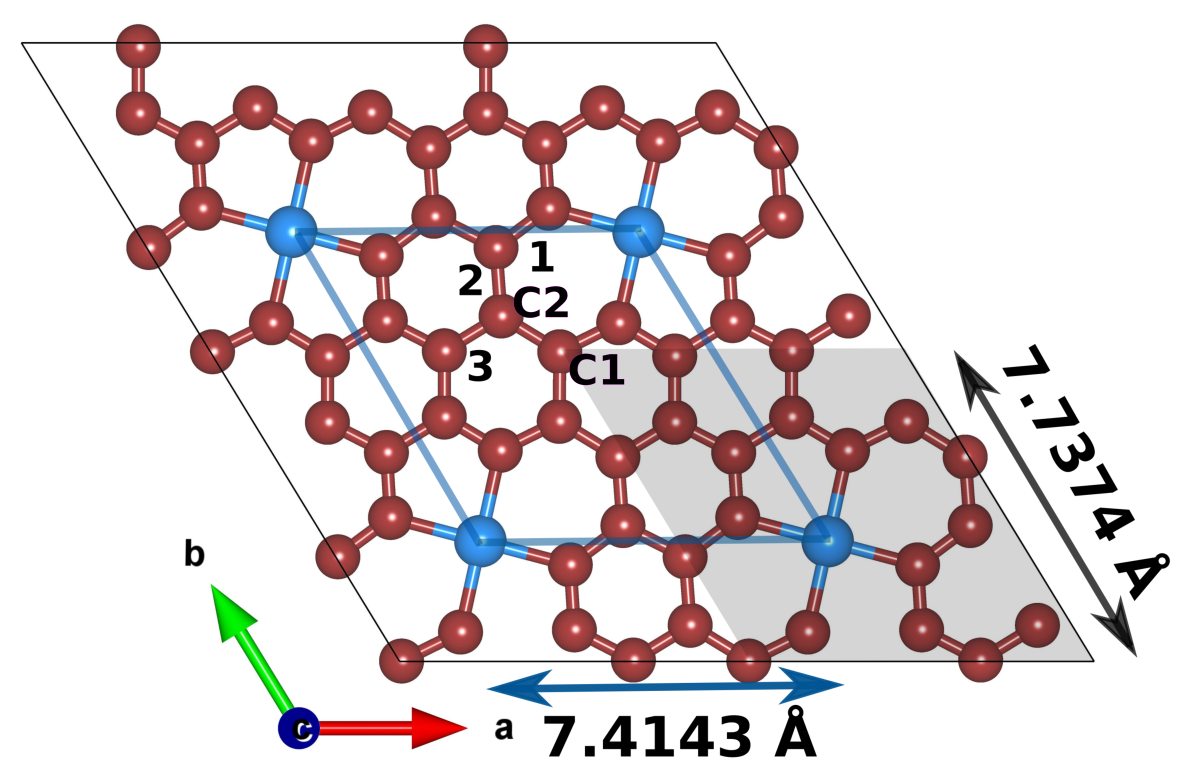}}
	\caption{Crystal structure of C$_{16}$Ge. Bond-lengths labelled by 1, 2, 3 are 1.4028\AA, 1.4660\AA and 1.4219 \AA, while the  C-Ge bond length is 1.9846\AA. C-Ge-C, C-C-Ge and C-C1-C, C-C2-C bond angles are 92.425$^o$, 128.508$^o$ and 122.502$^o$, 125.556$^o$.}
	\label{fig3}
\end{figure}

A C-C bond of 3$\times$3 supercell of graphene was replaced by a Ge atom to form the C$_{16}$Ge structure (see Fig. \ref{fig3}). The structure was relaxed using variable cell relaxation method. Ions were free to move in all direction. Due to divacancy formation after removal of a C-C bond, the space was adequate to host a Ge atom. No out of-the-plane buckling took place.

 The in-plane lattice constants are  7.4143\AA ~along lattice vector "a" and 7.7374\AA ~along lattice vector "b". C-Ge bond length is 1.8388\AA, whereas, the C-C bonds denoted by 1, 2, 3 in Fig.\ref{fig3} have lengths of  1.4028, 1.4660, 1.4219 \AA ~respectively. Here also, the hexagonal rings with only carbon atoms are not regular-hexagons. In addition, pentagonal rings with one Ge and four C atoms are formed. C-Ge-C, C-C-Ge bond angles are 92.425$^o$, 128.508$^o$ and there are two different types of C-C-C bond angle of 122.502$^o$, 125.556$^o$.

{\bf Electronic Properties:}\hfill\\

\begin{figure}[b]
	\centering
	\framebox{\includegraphics[scale=0.25]{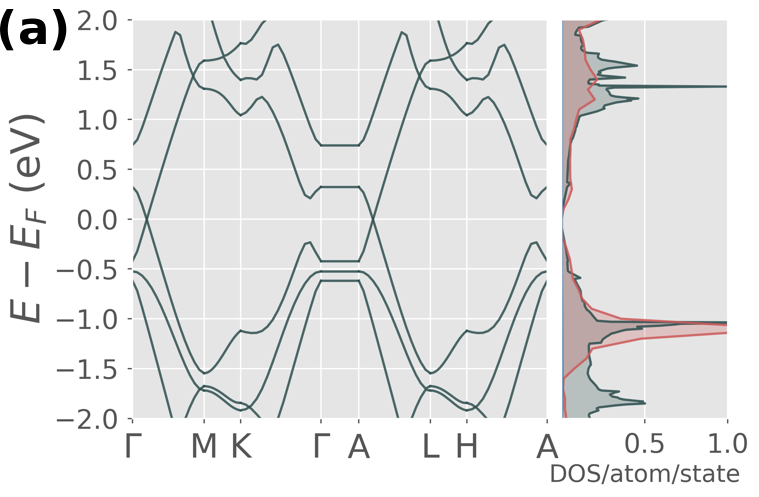}}
	\framebox{\includegraphics[scale=0.1]{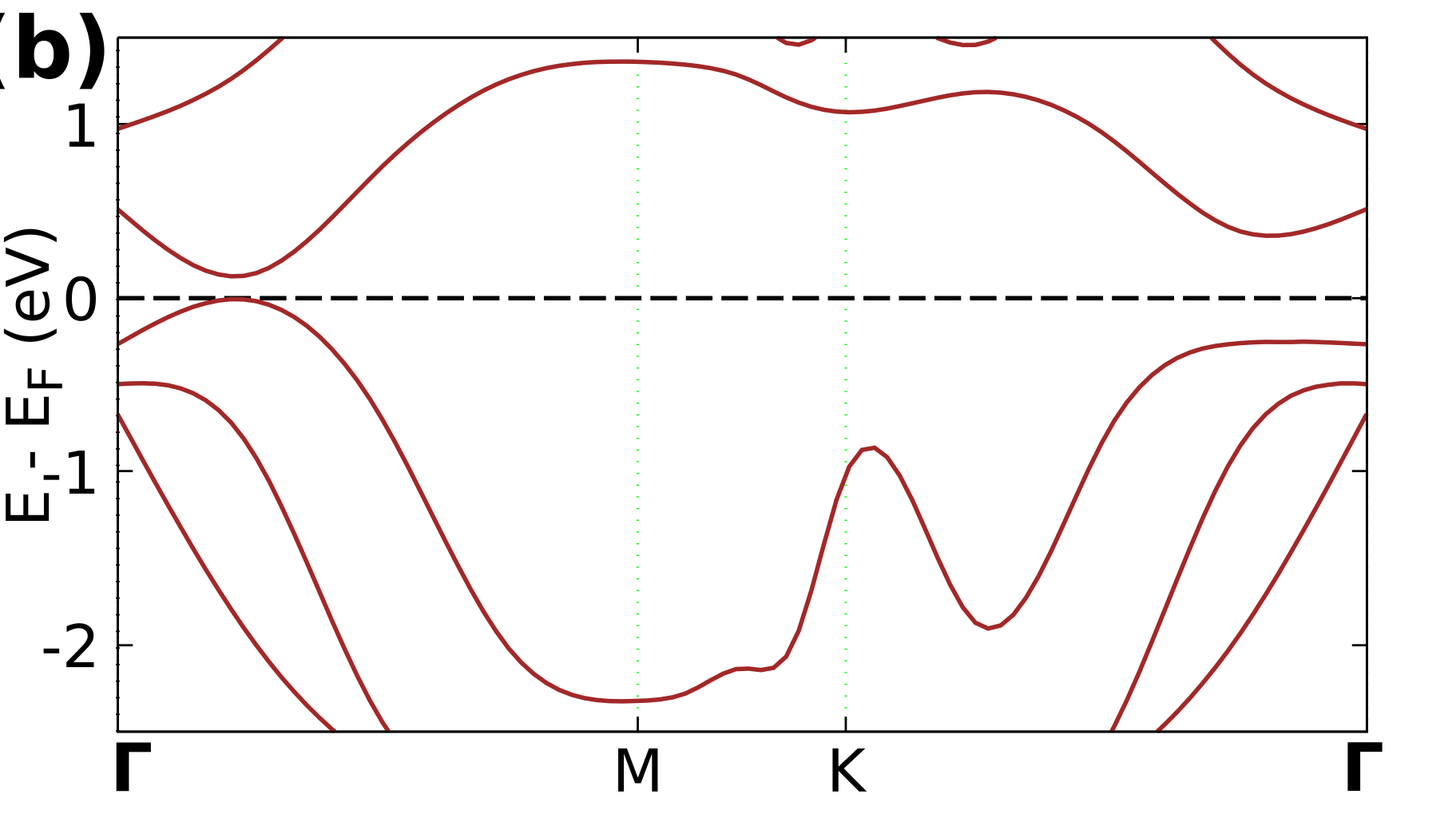}}
	\caption{(a)Band structure and density of state (DOS) plot about Fermi energy (E$_F$) using PBE for C$_{16}$Ge. The p$_z$ projected, p$_x$ projected DOS of Ge and total DOS are in red, blue and grey, respectively. (b) Band structure plot using HSE. PBE estimates it as a zero-gap material while HSE calculated gap is 0.1921 eV.)  }
	\label{fig4}
\end{figure}

In Fig. \ref{fig4}, we plot the band structure with respect to the Fermi-energy (E$_{F}$). The electronic band-structures and density of states (DOS) were calculated at the optimized lattice constants using PBE exchange-correlation. According to this plot, C$_{16}$Ge shows a metallic character. At E$_F$ band crossing takes place on the $\Gamma$ to M and A to L lines. The t-DOS plot has no sharp edge at either side of Fermi energy. The HO and LU states are dominated by Ge p$_z$ state (pDOS in red).

 Though PBE has predicted a metallic behaviour of C$_{16}$Ge, HSE hybrid functional calculation show a narrow band-gap of $0.1921 eV$. This is an interesting result. A semiconductor, even with a narrow gap is fundamentally very different from a metal. In Fig.\ref{fig4}(b), HSE calculated bands are presented. We can see that the gap opens up in the $\Gamma - M$ path. Like C$_{17}$Ge, this structure shows direct gap as well.

\begin{table}[h!]
	\vskip 0.1cm
	\centering
	\caption{Band Gap of C$_{17}$Ge and C$_{16}$Ge}
	\label{tab1}
	\begin{tabular}{c|cc}
		\hline \hline
		\multirow{2}{*}{\textbf{Structure}} & \multicolumn{2}{c}{\textbf{Band Gap (eV)}} \\ \cline{2-3} 
		& \textbf{PBE}         & \textbf{HSE}        \\ \hline
		\textbf{C$_{17}$Ge}                 &   0.2186             &   1.2271             \\
		\textbf{C$_{16}$Ge}                 &   Metallic             &   0.1921              \\ \hline
	\end{tabular}
\end{table}

\subsection{Optical Properties}

 Solar energy output is shared by: 5\% ultraviolet,45\% visible, and 50\% of infrared rays.\cite{36_new} So, the visible to near-infrared range offers the best opportunity to utilize solar energy. For absorption in that range, a band-gap of around 1 to 3.5 eV is suitable. The C$_{17}$Ge structure satisfies this condition. Furthermore, the direct nature of band-gap provides efficient photons to electron-hole pair conversion and nullifies the chance of energy loss during this process. The optimal band gap in C$_{17}$Ge brings forward its applicability as opto-electronic material.\cite{obg2}

\begin{figure}[h]
	\centering{
		\includegraphics[scale=0.35]{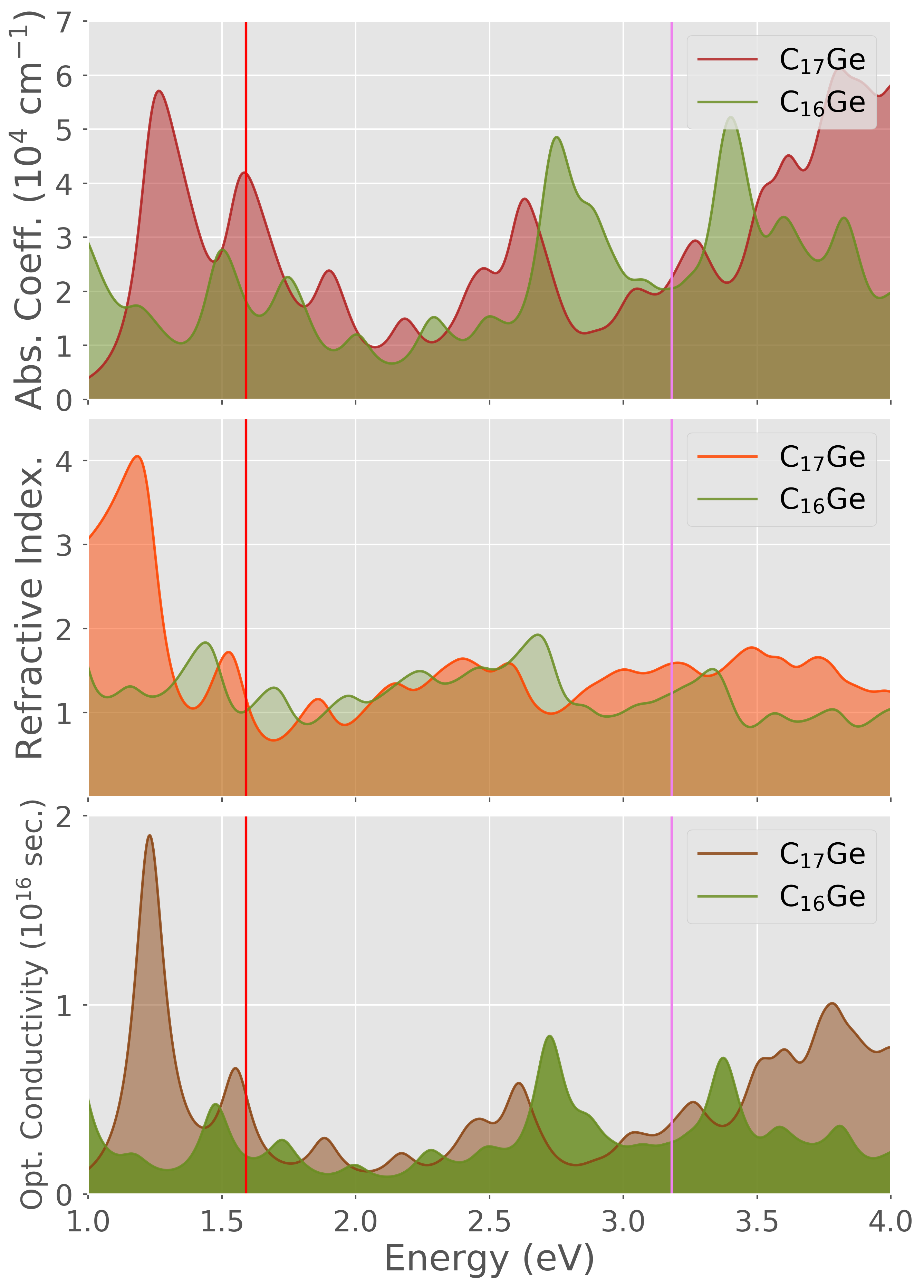} 
		\caption{ Graphical (a) absorption spectra, (b) refractive index, and (c) optical conductivity  as a function of photon energy calculated using hybrid functional (HSE).  
			\label{fig5}}
	}
\end{figure}

In Fig.~\ref{fig5}, we plot absorption coefficient, refractive index and  optical conductivity as a function of photon energy. Though our region of interest is visible to near infrared range, i.e., $\sim$ 1 to 4 eV, the  plot of effective electron number (n$_{eff}$) participating in inter-band transition justifies (0$-$15 eV) range. \cite{saha2000}: 

\begin{equation}
n_{eff}(E_m) = \frac{2m}{Ne^2h^2}\int_0^{E_m}E . \epsilon^{(i)}_{\alpha\alpha}(E) dE;
\end{equation}

In the visible to near infrared region, the absorption edge extents of the germagraphene structures are large enough to absorb the sunlight effectively. Band gaps of C$_{17}$Ge and C$_{16}$Ge are 1.2271 and 0.1921 eV, correspond to $1010.4 nm$ and $6454 nm$, respectively. So, depending on the band-gap value, C$_{17}$Ge is suitable for optoelectronic device production. The onset of absorption for C$_{17}$Ge is near 1.2 eV, which is very close to its band-gap. Interestingly, both of the structures show enhanced absorption in the range $4 - 6 eV$ (see, Fig. \ref{figA1}).

The refractive index of germanene structures are depicted as function of photon energy in Fig.~\ref{fig5}(b). In the visible range, the refractive index for C$_{16}$Ge varies from 1 to 1.9 and C$_{17}$Ge shows a mixed character by a variance in the range 0.8 to 1.6. There is a direct connection of optical absorption spectra with the imaginary part of the refractive index through the Eq. (4). Therefore, the nature of refractive index plots are expected to replicate the peaks and valleys of absorption spectra.
 
%Along with LDA+vLB potential \cite{singh2013, sujoy2019}, the HSE also provides accurate description of the electronic band structure. Thus, we believe that the present results of ban gaps and optical properties are reliable. 

The optical conductivities are compared in Fig.~\ref{fig5}(c). 
The conductivity starts with a gap, which points towards the semiconducting character of the structures. Below their respective band gap values, the optical conductivity is zero. Throughout the visible range the conductivity is almost flat. There is a sharp peak at $\sim 1.2 eV$ for C$_{17}$Ge corresponds to its band-gap value. Such sharp peak indicates that for a very small range of electronic spectra the structure is optically very active. This wavelength selection type nature or mono-chromaticity is useful for many optical applications like LASER.

\section{Conclusion}

 We show that the band-gap opens up due to Ge doping in graphene, specially for the C$_{17}$Ge and C$_{16}$Ge structures which were experimentally prepared by Tripathi \etal. \cite{expt-germagraphene} C$_{17}$Ge shows a direct gap of 1.2271 eV. The band gap of C$_{17}$Ge is suitable for opto-electronic device making. The optical property study also supports this claim. C$_{17}$Ge affirms better thermoelectric performance than graphene. In a sum up, we can conclude that the efficiency of germagraphene in futuristic applications is quite promising. 

%\newpage

\bibliography{germagraphene}

\appendix
\counterwithin{figure}{section}

\section{Figures} 
\begin{figure}[h]
	\centering
	\includegraphics[scale=0.2]{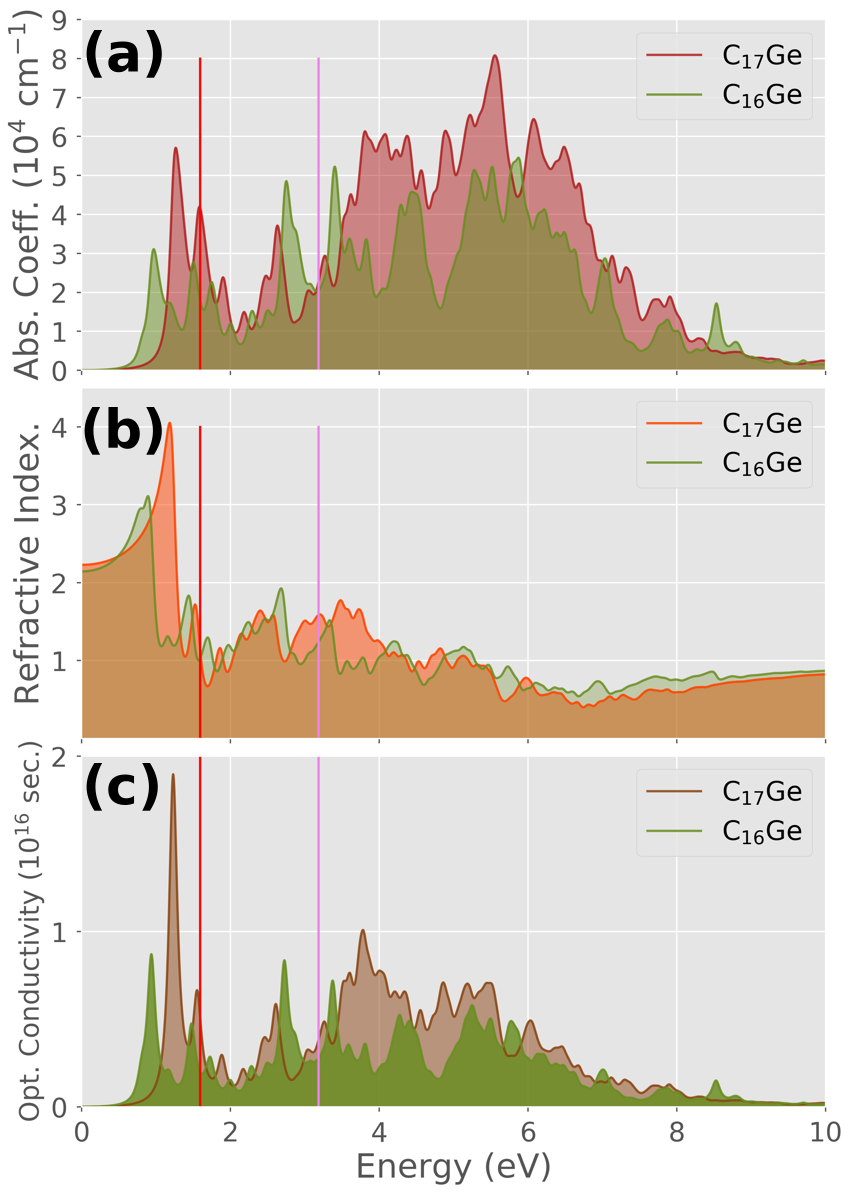}
	\caption{ Graphical (a) absorption spectra, (b) refractive index, and (c) optical conductivity  as a function of photon energy calculated using hybrid functional (HSE).  }
	\label{figA1}
\end{figure}
\end{document}